\documentclass[aip,pop,twocolumn, 10pt, floatfix]{revtex4-1} 
\usepackage{amsmath}
\usepackage{amssymb}
\usepackage{epsfig}
\usepackage{bm}

\newcommand{\vs}{\textit{vs.}\ }
\begin{document} 
\title{Gyrokinetic studies of the effect of $\beta$ on drift-wave stability in NCSX} 

\renewcommand{\thefootnote}{\alph{footnote}}

\author{J. A. Baumgaertel}%
\affiliation{%
Los Alamos National Laboratory, Los Alamos, New Mexico 87544
}%
\author{G. W. Hammett}
\affiliation{%
 Princeton Plasma Physics Laboratory, Princeton, New Jersey 08543
}%
\author{D. R. Mikkelsen}
\affiliation{%
 Princeton Plasma Physics Laboratory, Princeton, New Jersey 08543
}%
\author{M. Nunami}
\affiliation{
National Institute for Fusion Sciences, Japan
}
\author{P. Xanthopoulos}
\affiliation{%
 Max-Planck-Institut f\"{u}r Plasmaphysik, EURATOM Association, Teilinstitut Greifswald, Wendelsteinstr. 1, 17491 Greifswald, Germany}%

\date{\today}

\begin{abstract} 

 The gyrokinetic turbulence code GS2 was used to investigate the effects of plasma $\beta$ on linear, collisionless ion temperature gradient (ITG) modes and trapped electron modes (TEM) in National Compact Stellarator Experiment (NCSX) geometry. Plasma $\beta$ affects stability in two ways: through the equilibrium and through magnetic fluctuations. The first was studied here by comparing ITG and TEM stability in two NCSX equilibria of differing $\beta$ values, revealing that the high $\beta$ equilibrium was marginally more stable than the low $\beta$ equilibrium in the adiabatic-electron ITG mode case. However, the high $\beta$ case had a lower kinetic-electron ITG mode critical gradient. Electrostatic and electromagnetic ITG and TEM mode growth rate dependencies on temperature gradient and density gradient were qualitatively similar. The second $\beta$ effect is demonstrated via electromagnetic ITG growth rates' dependency on GS2's $\beta$ input parameter. A linear benchmark with gyrokinetic codes GENE and GKV-X is also presented.
\end{abstract}
\maketitle 
\section{Introduction}

Magnetic fusion energy requires the containment of very hot plasmas for a long enough time to allow fusion reactions to occur. Turbulent transport (most likely the result of drift-wave instabilities) breaks this confinement and can cause a significant amount of heat loss in tokamaks and spherical tori.\cite{liewer_measurements_1985} In contrast, neoclassical transport can often account for the poor confinement in traditional stellarators.\cite{fu_ideal_2007} However, modern stellarator designs, such as Wendelstein 7-AS (W7-AS),\cite{sapper_stellarator_1990} Wendelstein 7-X (W7-X),\cite{beidler_physics_1990,grieger_physics_1992} the National Compact Stellarator Experiment (NCSX),\cite{zarnstorff_physics_2001} the Large Helical Device (LHD),\cite{yamada_configuration_2001} and the Helically Symmetric Experiment (HSX) \cite{gerhardt_experimental_2005,canik_experimental_2007,talmadge_experimental_2008} have shown or are designed to have improved neoclassical confinement and stability properties. Thus, plasma turbulence and transport levels may be experimentally-relevant now and could affect performance of these stellarators. 

Gyrokinetic studies of drift-wave-driven turbulence in stellarator geometry are relatively recent and comprehensive scans are scarce. Most of these studies were done using upgraded versions of well-established axisymmetric codes: the linear eigenvalue FULL code,\cite{rewoldt_electromagnetic_1982,rewoldt_collisional_1987,rewoldt_drift_1999} the nonlinear initial-value or eigenvalue GENE code,\cite{jenko_electron_2000, xanthopoulos_nonlinear_2007} and the nonlinear initial-value code used in this paper, GS2. The nonlinear initial-value GKV-X code,\cite{watanabe_gyrokinetic_2007,nunami_gyrokinetic_2010} which uses the adiabatic electron approximation, was specifically written to simulate turbulence in stellarator geometry. All four codes use the flux-tube limit in their geometry, although GENE has been upgraded to allow for full flux-surface simulations. The microinstability code GS2\cite{dorland_electron_2000} was extended from its original axisymmetric-geometry version to treat the more general case of non-axisymmetric stellarator geometry as described by Ref. \onlinecite{baumgaertel_simulating_2011} and briefly mentioned in Section \ref{sec:coord}. The work in Ref. \onlinecite{baumgaertel_simulating_2011} includes a linear benchmark of GS2 stellarator simulations with FULL. Section \ref{sec:bench} briefly displays the results of a linear benchmark of GS2 with GENE and GKV-X, rounding out the major stellarator codes.

Finally, in Section \ref{sec:betaStudies} the upgraded GS2 is used for comprehensive parameter scans and instability studies in the National Compact Stellarator Experiment (NCSX) design. NCSX, with its quasi-axisymmetric magnetic configuration, is a bridge in configuration space between tokamaks and the rest of the stellarator world. Therefore, it is an excellent configuration to begin detailed gyrokinetic stellarator studies with GS2, which has been used successfully on axisymmetric geometry for many years. The two NCSX configurations used in Section \ref{sec:betaStudies} were created as part of a series of flexibility studies \cite{pomphrey_ncsx_2007} that were performed using a magnetic coil set similar to the final design of the NCSX machine. The equilibrium optimization code STELLOPT was used to find currents in these coils needed to meet desired configuration properties. As a consequence, sets of configurations exist in which only one parameter, such as magnetic shear and plasma $\beta$, varies significantly. Studies in Section \ref{sec:betaStudies} survey linear stability in two configurations that differ only by plasma equilibrium $\beta$. These equilibria were compared via the growth rates of the electrostatic adiabatic ITG mode, electrostatic collisionless kinetic ITG-TEM mode, and electromagnetic collisionless kinetic ITG-TEM mode.

\section{GS2 coordinate system}\label{sec:coord}

First, GS2 geometry input must be built by a series of programs. VMEC\cite{hirshman_momcon:_1986,hirshman_improved_1990} creates the 3D MHD equilibria used as a basis for all gyrokinetic stellarator codes. GIST\cite{xanthopoulos_geometry_2009} extracts from the full 3D equilibrium the geometrical data needed to represent a flux tube based on field-line-following coordinates.  This coordinate system includes the radial coordinate, $\rho=\sqrt{s}$ ($s\approx (r/a)^2$ is the normalized toroidal flux), the distance along a field line, $\theta$, and the angle that selects a flux tube, $\alpha=\zeta-q(\theta-\theta_0)$ (where $\zeta$ and $\theta$ are the Boozer toroidal and Boozer poloidal coordinates and $\theta_0$ is the ballooning parameter). To obtain the final GS2 geometry input file, FIGG\cite{baumgaertel_simulating_2012} uses the GIST output file to calculate the pitch angle parameter grid, involved in the velocity integration of the distribution function. (The pitch angle parameter, $\lambda=\mu/E$, is related to $v_{||}$ through $v_{||}/v=\sqrt{1-\lambda B}$.) 

The GS2 documentation\cite{barnes_trinity:_2009} defines geometrical quantities in terms of a parameter $d\Psi_N/d\rho$,  where $\rho$ is the radial coordinate and $\Psi_N$ is the normalized poloidal flux. Geometrical quantities in this paper follow GS2 notation and include $d\Psi_N/d\rho$. For more information, see Refs. \onlinecite{baumgaertel_simulating_2011,baumgaertel_simulating_2012}.

\section{Benchmarks of GS2, GENE, GKV-X}\label{sec:bench}

GS2, GENE, and GKV-X results were compared for an NCSX VMEC equilibrium based on the standard S3 configuration of NCSX design. This configuration is quasi-axisymmetric with three field periods, an aspect ratio of 3.5, and a major radius of 1.4 m. The following benchmark used geometry with the surface at $s=0.5$ ($r/a\approx0.7$), the $\alpha=0$ field line, and the ballooning parameter $\theta_{0}=0$. The average $\beta$ is $\langle\beta\rangle=4\%$ and, at this surface, the safety factor is $q=1.978$. 

\begin{figure} \centering
\includegraphics[scale=1.0]{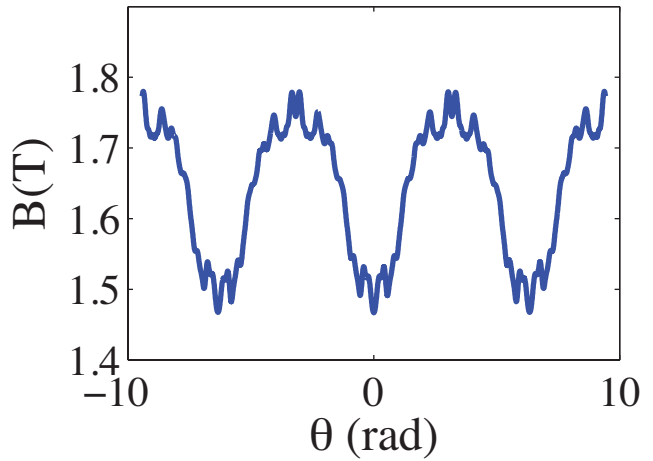}
\caption{B \vs $\theta$ grid for NCSX QAS3, with $s=0.5$, $\alpha=0$,
and $\theta_{0}=0$. (color online)}
\label{fig:geneNCSXbmag}
\end{figure}

Figure \ref{fig:geneNCSXbmag} shows the variation of the magnitude of the magnetic field along the chosen magnetic field line, with a resolution of 209 $\theta$ grid points per poloidal period. There were approximately 30 points in the pitch angle parameter grid. The $\theta$ range extends from $-3\pi$ to $3\pi$.

\begin{figure} \centering
\includegraphics[scale=1.0]{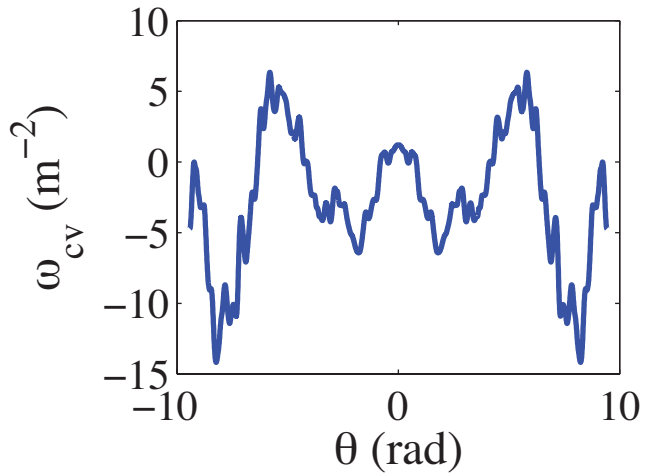}
\caption{Variation of the curvature drift term ($\omega_{cv}=(d\Psi_N/d\rho)(\mathbf{k_\perp}/n)\cdot \mathbf{b}\times [\mathbf{b} \cdot \nabla \mathbf{b}]$) (for $n=1$) along $\theta$ for NCSX QAS3, with $s=0.5$, $\alpha=0$, and $\theta_{0}=0$. (color online)}
\label{fig:geneNCSXcurvdr}
\end{figure}

\begin{figure} \centering
\centering
\includegraphics[scale=1.0]{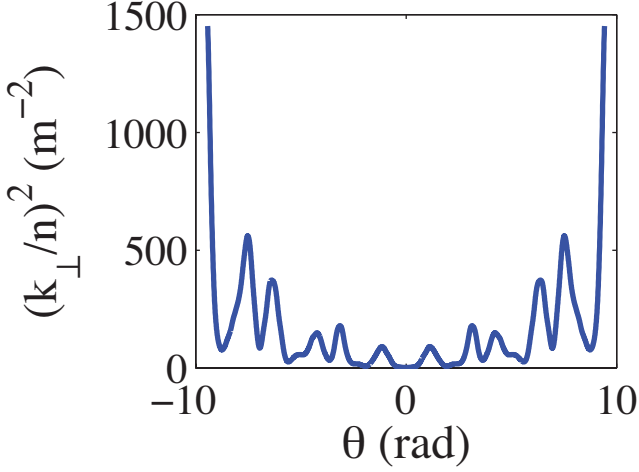}
\caption{Variation of $(\frac{k_{\perp}}{n})^{2}(\theta)(d\Psi_N/d\rho)^2$ for NCSX QAS3, with $s=0.5$, $\alpha=0$, and $\theta_{0}=0$. (color online)}
\label{fig:geneNCSXgds2}
\end{figure}

Figures \ref{fig:geneNCSXcurvdr} and \ref{fig:geneNCSXgds2} are the variations of $(k_{\perp}/n)^{2}$, where $n$ is the toroidal mode number, and the curvature drift along the same chosen field line. By convention, positive curvature drifts are ÒbadÓ or destabilizing, while negative curvature drifts are ÒgoodÓ or stabilizing. Significant unstable modes occur where $k_\perp$ is small, which is near $\theta=0$ for this equilibrium, since instabilities are generally suppressed at large $k_\perp$ by FLR averaging. Because Figure \ref{fig:geneNCSXcurvdr} indicates that the curvature is bad in this region near $\theta=0$, where $k_\perp$ is the smallest, it is expected that unstable modes will appear here.

The benchmark case was an electrostatic, collisionless ITG mode with adiabatic electrons. The temperatures were such that $T_i=T_e$, the temperature gradient was $a_N/L_{T}= 3$ ($a_N/L_{X}=-a_N(1/X)dX/d\rho$), and the density gradient was $a_N/L_{n}= 0$, where the normalizing length was chosen to be an averaged minor radius, $a_N\approx 0.323m$. See Table \ref{tab:geneNCSXbench}.

\begin{figure} \centering
\includegraphics[scale=1.0]{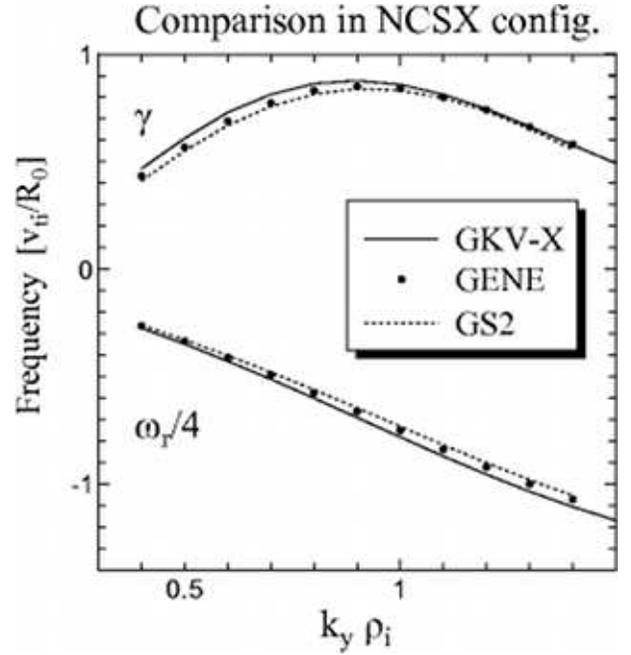}
\caption{Variation of $\gamma$ and $\omega_r$ with $k_{y}\rho_{i}$ for NCSX QAS3, comparing three codes.}
\label{fig:gs2genegkvxNCSX}
\end{figure}

\begin{figure} \centering
\includegraphics[scale=1.0]{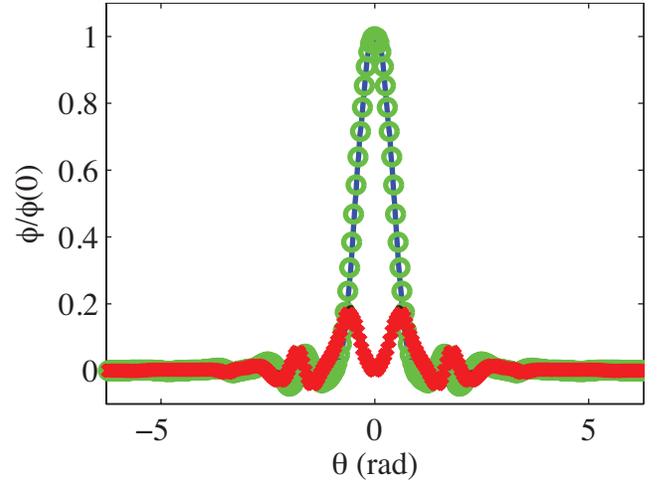}
\caption{Comparison of $\phi$ \vs $\theta$ (with $k_y\rho_i=0.9$) for GS2 ($Re(\phi)$: blue line, $Im(\phi)$: black line) and GKV-X ($Re(\phi)$: green circles, $Im(\phi)$: red crosses). (color online)}
\label{fig:genegkvxNCSXefcn}
\end{figure}

Figure \ref{fig:gs2genegkvxNCSX} shows the growth rate and real frequency spectra for this mode. The maximum discrepancy in growth rate between GS2 and GKV-X is $8\%$, with GENE always in between. The frequencies agree to within $5\%$. This agreement is excellent. GS2 and GKV-X's electrostatic potential, $\phi$, for a particular $k_y\rho_i=0.9$ is shown in Figure \ref{fig:genegkvxNCSXefcn}. These electrostatic eigenfunctions also agree well. 

\begin{table} \centering
\begin{tabular}{|c|c|}
\hline 
$s \approx \left(\langle r/a \rangle \right)^2$ & $0.5$\tabularnewline
\hline 
$\alpha=\zeta-q\theta$  & $0$\tabularnewline
\hline 
$\theta_{0}$  & $0$\tabularnewline
\hline 
$q$ & $1.978$\tabularnewline
\hline 
$\langle\beta\rangle$ & $4\%$\tabularnewline
\hline 
$T_{i}=T_{e}$ & $1\mathrm{keV}$\tabularnewline
\hline 
$a_{N}/L_{ni}=a_{N}/L_{ne}$ & $0$\tabularnewline
\hline 
$a_{N}/L_{Ti}=a_{N}/L_{Te}$ & $3$\tabularnewline
\hline 
$R_0$ & $\approx4a_{N}\approx 1.4m$\tabularnewline
\hline 
$a_{N}$ & $\approx 0.323m$\tabularnewline
\hline
$B_{a}=\langle B \rangle$ & $\approx 1.6 T$ \tabularnewline
\hline
$m_{ref}$ & $2m_p$\tabularnewline
\hline
$v_t$ & $\sqrt{T_i/m_{ref}}$\tabularnewline
\hline
GS2 $\omega$ units $v_t/a_N$ & $\approx6.782\times 10^5 sec^{-1}$\tabularnewline
\hline
\end{tabular}
\caption{The set of local parameters used the microinstability
simulation based on the NCSX QAS3 equilibrium.}
\label{tab:geneNCSXbench}
\end{table}

Now, all four gyrokinetic stellarator codes have been benchmarked linearly (here GS2, GENE, and GKV-X; Ref. \onlinecite{baumgaertel_simulating_2011} describes GS2 and FULL's benchmark). In addition to those in NCSX geometry, comparisons of linear GENE and GS2 results for W7-X also agreed well.\cite{baumgaertel_simulating_2012} With these successful benchmarks, further studies can be performed with more confidence. The next section begins this venture with GS2.

\section{NCSX $\beta$ Studies}\label{sec:betaStudies}

High plasma $\beta$ is important for fusion because the fusion power at fixed magnetic field is approximately proportional to $\beta^2$. To start studying the effect of plasma beta on gyrokinetic turbulence in stellarators, linear ITG and TEM stability was compared for two configurations, one with equilibrium $\beta=0\%$ and one with $\beta=4\%$. Figure 6 of Ref. \onlinecite{pomphrey_ncsx_2007} shows the poloidal cross-sections for three toroidal locations, along with $\iota$ profiles for various plasma currents ($I_p$).  The plasma shape varies very little with $I_p$. This section uses set of beta scans with $I_p=174kA$, because their $\iota$ profiles varied the least, allowing for isolation of the effects of $\beta$. 

\subsection{Discussion of GS2 $\beta$ Parameter}

The physical beta enters into GS2's equations (the gyrokinetic and Maxwell's equations, see Ref. \onlinecite{gs2home_sourceforge.net:_????}) in two main ways, through its indirect effect on the MHD equilibrium (such as the Shafranov shift and the curvature drift) and through its direct effect in the gyrokinetic equations, controlled through the parameter $\beta_{input}$. This GS2 beta parameter is defined as $\beta_{input}=2\mu_0n_{ref}T_{ref}/B_{ref}^2$, the ratio of the reference pressure to the reference magnetic energy density. $\beta_{input}$ is used in the scaling of $\delta B_{||}=\nabla_{\perp}\times\mathbf{A}_{\perp}$ and $\delta A_{||}$, through, for example, the weighting of the contribution of each species to the total parallel current by a factor $w_s=2\beta_{input}Z_sn_s\sqrt{T_s/m_s}$. While $\beta_{input}$ must be set to match $\beta_{equil}$ in the geometry files for consistent results, setting $\beta_{input}=0$ is a convenient way to turn off magnetic fluctuations and focus only on electrostatic fluctuations, as done in Sections \ref{sec:adi}-\ref{sec:kin}.

\subsection{Geometry and Plasma Parameters}\label{sec:ncsxFlexGeo}

\begin{figure}
\centering
\includegraphics[scale=1.0]{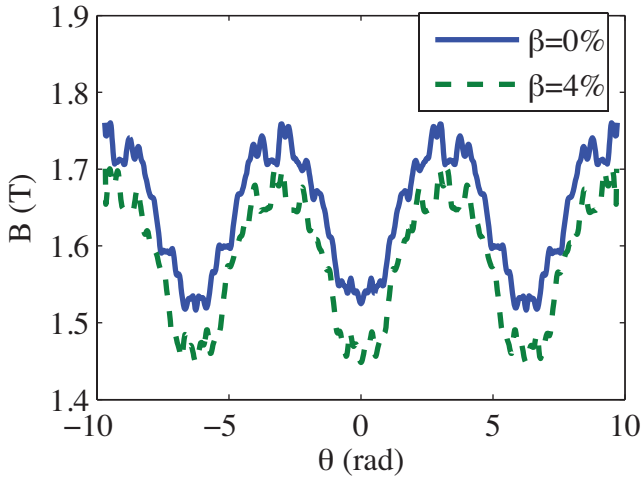}
\caption{NCSX beta flexibility studies comparing $|B|$ \vs $\theta$ for both $\beta=0\%$ and $\beta=4\%$, at $\sqrt{s}=r/a=0.5$, $\alpha=0$, and $\theta_0=0$. (color online)}
\label{fig:ncsxflexBMAG}
\end{figure}

\begin{figure}
\centering
\includegraphics[scale=1.0]{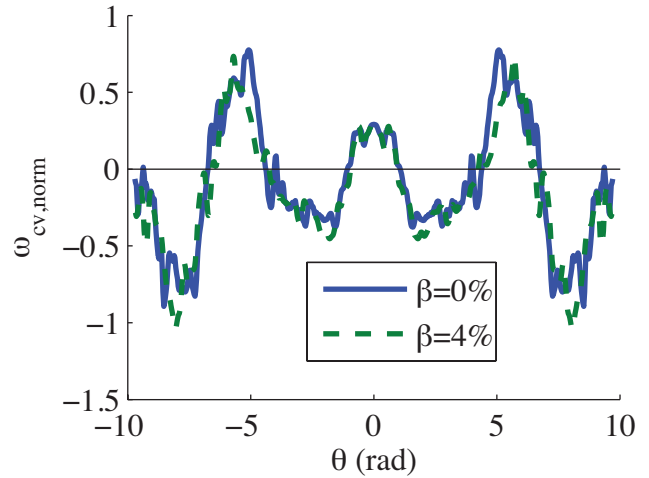}
\caption{NCSX beta flexibility studies comparing the curvature drift terms ($\omega_{cv,norm}=(2a_N^2/B_N)(d\Psi_N/d\rho)(k_\perp/n)\cdot \mathbf{b}\times [\mathbf{b}\cdot\nabla \mathbf{b}]$) along $\theta$, for $\beta=0\%$ and $4\%$, at $\sqrt{s}=r/a=0.5$, $\alpha=0$, and $\theta_0=0$. (color online)}
\label{fig:ncsxflexDRIFTS}
\end{figure}

\begin{figure}
\centering
\includegraphics[scale=1.0]{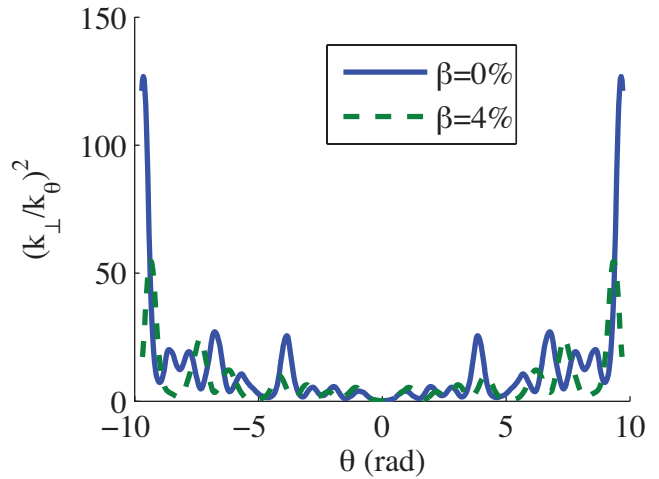}
\caption{NCSX beta flexibility studies comparing $\left(\frac{k_\perp}{k_\theta}\right)^2$ \vs $\theta$ for $\beta=0\%$ and $4\%$, at $\sqrt{s}=r/a=0.5$, $\alpha=0$, and $\theta_0=0$. (color online)}
\label{fig:ncsxflexKPERP}
\end{figure}

For these $\beta$ studies, the geometry used had the surface with normalized toroidal flux $s\approx\left(\langle r/a\rangle\right)^2=0.26$, field line $\alpha=0$, and ballooning parameter $\theta_0=0$. The magnitude of magnetic field, curvature and $\nabla B$ drift components, and $|k_\perp|^2$ along the field line are plotted for both $\beta=0\%$ and $\beta=4\%$ in Figures \ref{fig:ncsxflexBMAG}-\ref{fig:ncsxflexKPERP}.  More parameters for both equilibria are in Table \ref{tab:betaStudiesGeoBeta04}. All growth rate and frequency values are normalized such that $(\gamma,\omega)=(\gamma_{physical},\omega_{physical})(a/v_{thi})$. These runs are collisionless (collision frequency $\nu=0$). For the following studies, several plasma parameters were varied around the base case parameters shown in Table \ref{tab:betaStudiesParams}.

For each equilibrium ($\beta=0\%$ and $\beta=4\%$), convergence studies were run with increasing resolution in $\theta$ and velocity-space for single ion species, ITG-driven adiabatic electron modes. Here, GS2 studies use grids with $30$ pitch angle parameter points for both the $\beta=0\%$ and $4\%$ equilibria, approximately $750$ $\theta$ points for $\beta=4\%$, and approximately $630$ $\theta$ points for $\beta=0\%$. These grids' results were well-converged (to within a few percent of the results from higher-resolution grids).  There are $32$ energy grid points.%

\begin{table}
\centering
\begin{tabular}{|c|c|c|}
\hline 
 Parameter & $\beta=0\%$ & $\beta=4\%$\tabularnewline
\hline 
$s \approx \left(\langle r/a \rangle \right)^2$ & $0.26$ & $0.26$\tabularnewline
\hline 
$\alpha=\zeta-q\theta$  & $0$ & $0$\tabularnewline
\hline 
$\theta_{0}$  & $0$ & $0$\tabularnewline
\hline 
$q_s$ & $2.175$ & $2.011$\tabularnewline
\hline 
$\hat{s}$& $0.356$& $0.278$\tabularnewline
\hline 
$\langle\beta\rangle$ & $0.0\%$& $4\%$\tabularnewline
\hline 
$R$ & $\approx4.7a_{N}\approx1.5m$ &$\approx4.7a_{N}\approx1.5m$\tabularnewline
\hline 
$a_{N}$ & $\approx0.322m$& $\approx0.322m$\tabularnewline
\hline
$B_{a}=\langle B \rangle$ & $1.58 T$& $1.55 T$ \ \tabularnewline
\hline
\end{tabular}

\caption{Geometry values for the NCSX $\beta=0\%$ and $4\%$ equilibria.}
\label{tab:betaStudiesGeoBeta04}
\end{table}

\begin{table}
\centering
\begin{tabular}{|c|c|}
\hline 
$k_y\rho_i$ & $\in[0.6, 1.4]$\tabularnewline
\hline 
$T_{i}=T_{e}$ & $1keV$\tabularnewline
\hline 
$\nu$ & $0$\tabularnewline
\hline
$m_{ref}$ & $2m_p$\tabularnewline
\hline
$v_t$ & $\sqrt{T_i/m_{ref}}$\tabularnewline
\hline
GS2 $\omega$ units $v_t/a_N$ & $\approx6.214\times 10^5 sec^{-1}$\tabularnewline
\hline
\end{tabular}

\caption{The base set of local parameters used in the NCSX $\beta$ studies.}
\label{tab:betaStudiesParams}
\end{table}

\subsection{Electrostatic Adiabatic ITG mode}\label{sec:adi}
 
\begin{figure} \centering
\includegraphics[scale=1.0]{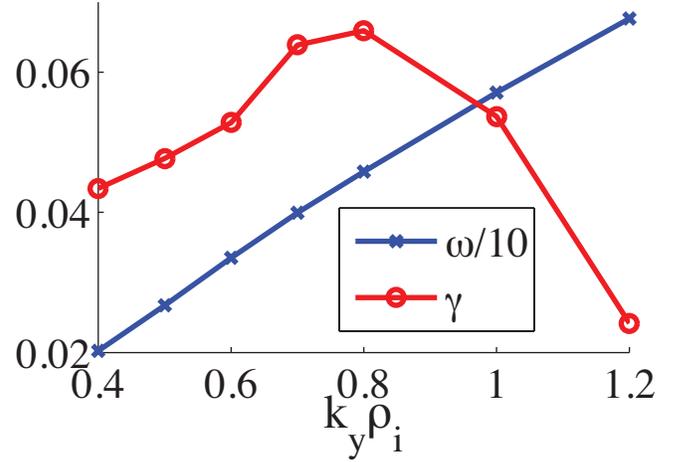}
\caption{ITG adiabatic electron growth rates \vs $k_y\rho_i$ for NCSX $\beta_{equil}=0\%$, $a_N/L_n=0$, and $a_N/L_T=1.6$. (color online)}
\label{fig:ncsxITGaetp16kyrho}
\end{figure}

\begin{figure} \centering
\includegraphics[scale=1.0]{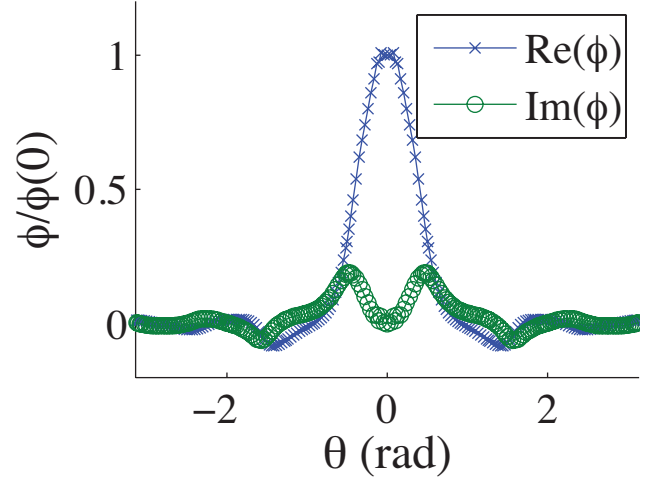}
\caption{A representative electrostatic potential eigenfunction for these ITG modes. $a_N/L_T=1.6$, $a_N/L_n=0$, and $k_y\rho_i=1$.}%
\label{fig:adiEfcn}
\end{figure}

\begin{figure} \centering
\includegraphics[scale=1.0]{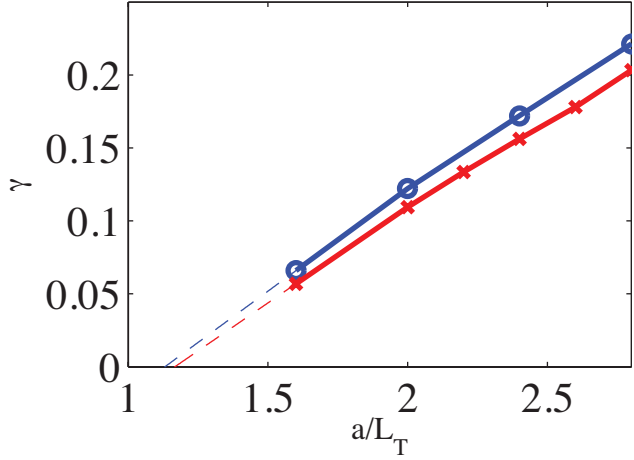}
\caption{ITG adiabatic electron growth rates \vs temperature gradient for NCSX $\beta_{equil}=0\%$ (blue circles) and $4\%$ (red crosses), $a_N/L_n=0$. Fits (dashed lines) obtained through piecewise linear interpolation on the lowest half of the growth rate curve. (color online)}
\label{fig:ncsxITGaecritgrad}
\end{figure}

Using the base parameters in Section \ref{sec:ncsxFlexGeo}, linear ITG stability as a function of temperature gradient, $a_N/L_T$, was compared for both equilibria, over a wavenumber range of $k_y\rho_i\in[0.6,1.4]$ (note in Figure \ref{fig:ncsxITGaetp16kyrho} that this range is sufficient to capture the peak of the growth rate spectrum, and therefore the fastest growing mode). The peak growth rates for both equilibria occur between $k_y\rho_i\approx0.6$ and $\approx 1.2$ and are shown in Figure \ref{fig:ncsxITGaecritgrad}, indicating that the critical gradient of the $\beta=0\%$ equilibrium is $a_N/L_{T,crit}\approx1.13$ and that of the  $\beta=4\%$ equilibrium is $a_N/L_{T,crit}\approx1.16$. These values are not significantly different. The fact that beyond marginal stability, the growth rates of $\beta=0\%$ are larger than those of $\beta=4\%$ is an indication that $\beta$ is stabilizing to the ITG mode, as found in the tokamak studies in Ref. \onlinecite{pueschel_transport_2010}. A representative eigenfunction is shown in Figure \ref{fig:adiEfcn}; note the typical ballooning-about-zero behavior of an ITG mode.

\begin{figure} \centering
\includegraphics[scale=1.0]{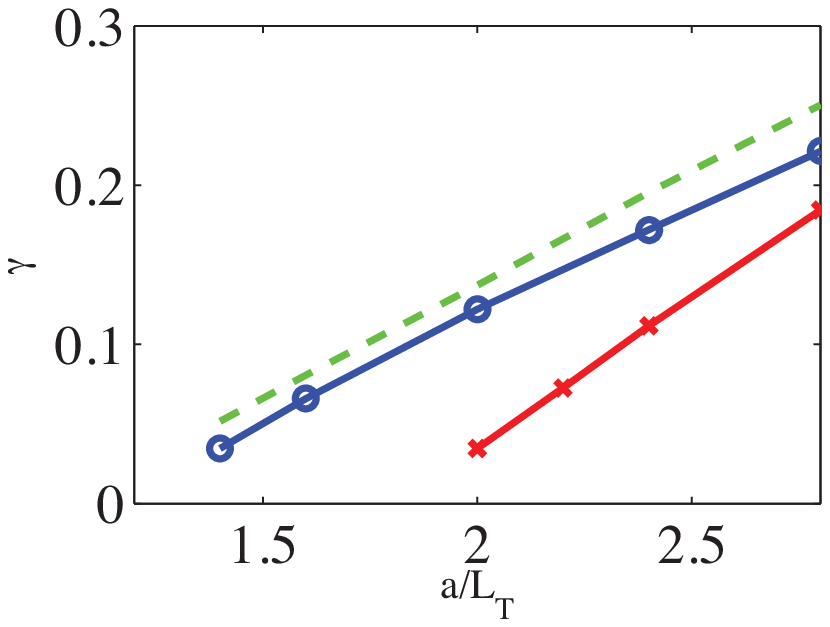}
\caption{Electrostatic ITG adiabatic electron growth rates (at $k_y\rho_i$ of maximum $\gamma$) \vs temperature gradient for NCSX $\beta_{equil}=0\%$ for various density gradients: $a_N/L_n=0$ (blue circles), $a_N/L_n=1$ (green dashed line), $a_N/L_n=2$ (red crosses). (color online)}
\label{fig:ncsxbeta0ITGaetprim_varifprim}
\end{figure}

\begin{figure} \centering
\includegraphics[scale=1.0]{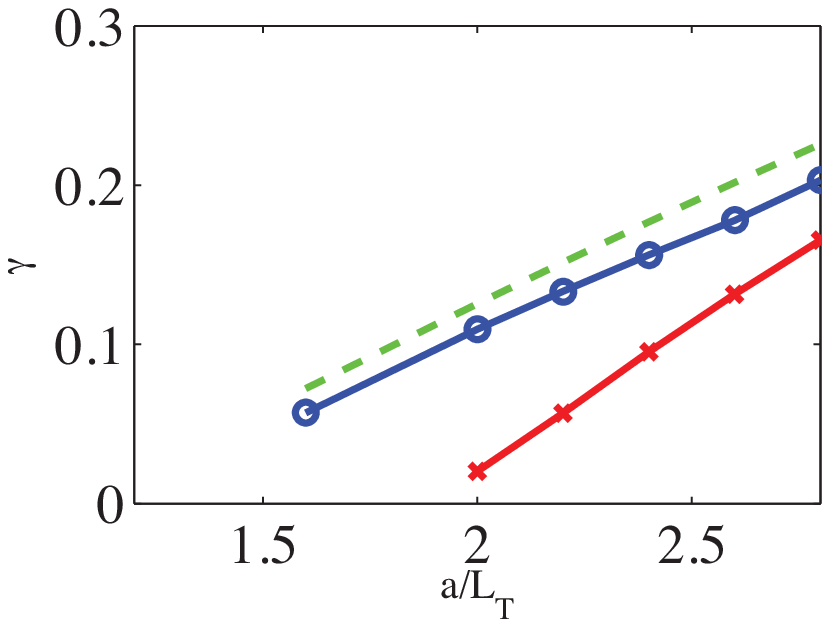}
\caption{Electrostatic ITG adiabatic electron growth rates (at $k_y\rho_i$ of maximum $\gamma$) \vs temperature gradient for NCSX $\beta_{equil}=4\%$ for various density gradients: $a_N/L_n=0$ (blue circles), $a_N/L_n=1$ (green dashed line), $a_N/L_n=2$ (red crosses). (color online)}
\label{fig:ncsxbeta4ITGaetprim_varifprim}
\end{figure}

Looking at the effect of the density gradient on the critical temperature gradient in Figures \ref{fig:ncsxbeta0ITGaetprim_varifprim}-\ref{fig:ncsxbeta4ITGaetprim_varifprim}, with $a_N/L_n=1$, $a_N/L_{T,crit}$ lowers by $\approx0.1$ with respect to the $a_N/L_n=0$ value in each case, appearing to be somewhat destabilizing.  $a_N/L_n\geq2$, however, appears to be strongly stabilizing, consistent with a transition to the slab limit of the ITG mode where a density gradient is stabilizing.\cite{jenko_critical_2001}

\begin{figure} \centering
\includegraphics[scale=1.0]{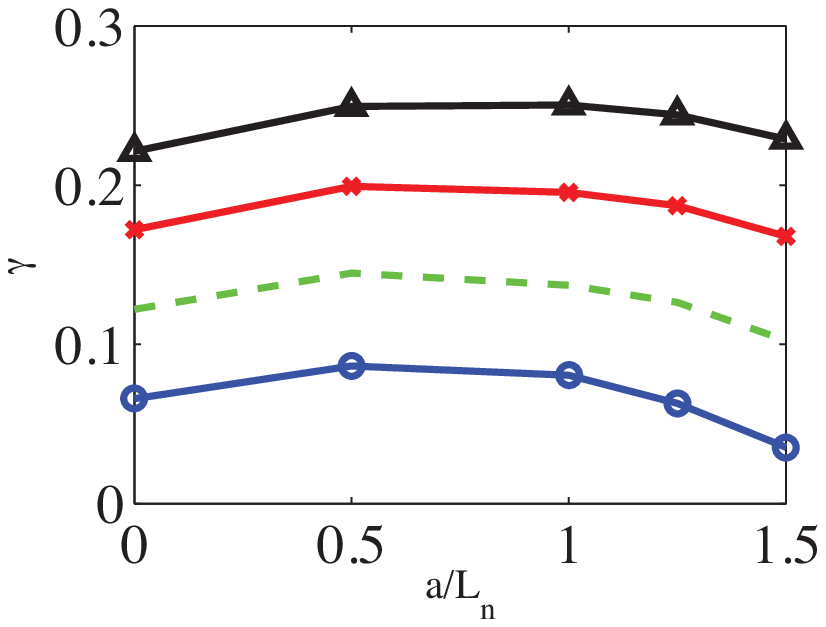}
\caption{Electrostatic ITG adiabatic electron growth rates (at $k_y\rho_i$ of maximum $\gamma$) \vs density gradient for NCSX $\beta_{equil}=0\%$ for various temperature gradients: $a_N/L_T=1.6$ (blue circles), $a_N/L_T=2.0$ (green dashed line), $a_N/L_T=2.4$ (red crosses), $a_N/L_T=2.8$ (black triangles). (color online)}
\label{fig:ncsxbeta0ITGaefprim_varitprim}
\end{figure}

\begin{figure} \centering
\includegraphics[scale=1.0]{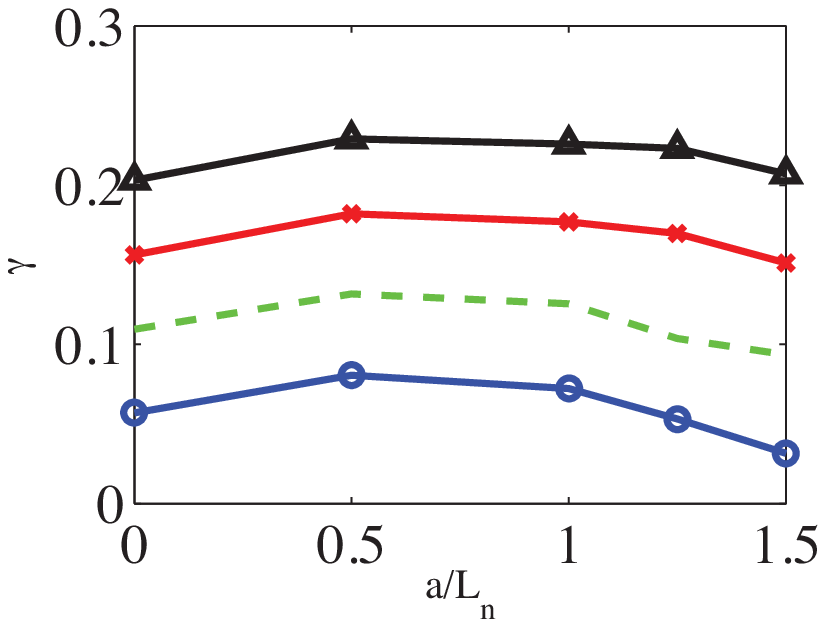}
\caption{Electrostatic ITG adiabatic electron growth rates (at $k_y\rho_i$ of maximum $\gamma$) \vs density gradient for NCSX $\beta_{equil}=4\%$ for various temperature gradients: $a_N/L_T=1.6$ (blue circles), $a_N/L_T=2.0$ (green dashed line), $a_N/L_T=2.4$ (red crosses), $a_N/L_T=2.8$ (black triangles). (color online)}
\label{fig:ncsxbeta4ITGaefprim_varitprim}
\end{figure}

As expected, when one compares the growth rates as a function of $a_N/L_n$ for various values of $a_N/L_T$ (Figs.  \ref{fig:ncsxbeta0ITGaefprim_varitprim}-\ref{fig:ncsxbeta4ITGaefprim_varitprim}), the growth rates increase monotonically with $a_N/L_T$. Also, the growth rates for $\beta=0\%$ are higher than those for $\beta=4\%$, another sign that higher $\beta$ could be somewhat stabilizing.

\subsection{Electrostatic Kinetic ITG-TEM}\label{sec:kin}

\begin{figure} \centering
\includegraphics[scale=1.0]{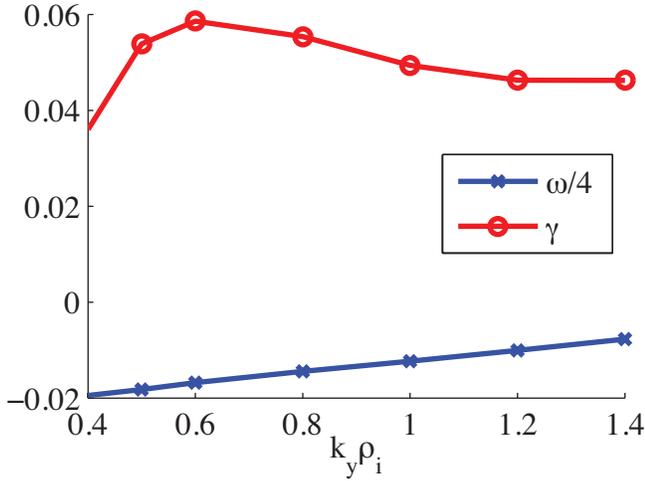}
\caption{Electrostatic ITG-TEM kinetic electron growth rates \vs $k_y\rho_i$ for NCSX $\beta_{equil}=0\%$, $a_N/L_n=1$, $a_N/L_T=0.5$. (color online)}
\label{fig:ncsxbeta0ITGkinetp05kyrho}
\end{figure}

\begin{figure} \centering
\includegraphics[scale=1.0]{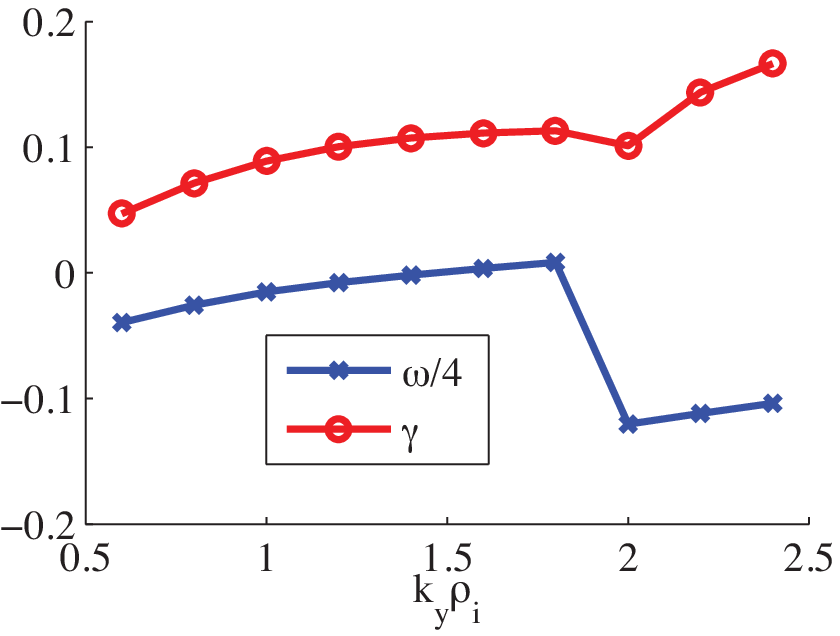}
\caption{Electrostatic ITG-TEM kinetic electron growth rates \vs $k_y\rho_i$ for NCSX $\beta_{equil}=0\%$, $a_N/L_n=2$, $a_N/L_T=0$. (color online)}
\label{fig:ncsxbeta0ITGkinetp2kyrho}
\end{figure}

Adding kinetically-treated electrons allows one to study the trapped electron mode (TEM) and hybrid ITG-TEM (driven by both $a_N/L_T$ and $a_N/L_n$) modes. Figure \ref{fig:ncsxbeta0ITGkinetp05kyrho} shows the $k_y\rho_i$ spectrum for $a_N/L_n=1,a_N/L_T=0.5$ and  Figure \ref{fig:ncsxbeta0ITGkinetp2kyrho} the $k_y\rho_i$ spectrum for $a_N/L_n=2,a_N/L_T=0$. The peak of the growth rate spectrum shifts as $a_N/L_n$ and $a_N/L_T$ change. When the gradients are large enough (Figure \ref{fig:ncsxbeta0ITGkinetp2kyrho}), two distinct regimes are seen, with a mode switch evident in the frequencies. These studies focus on the peaks lower than $k_y\rho_i=1.8$.

\begin{figure} \centering
\includegraphics[scale=1.0]{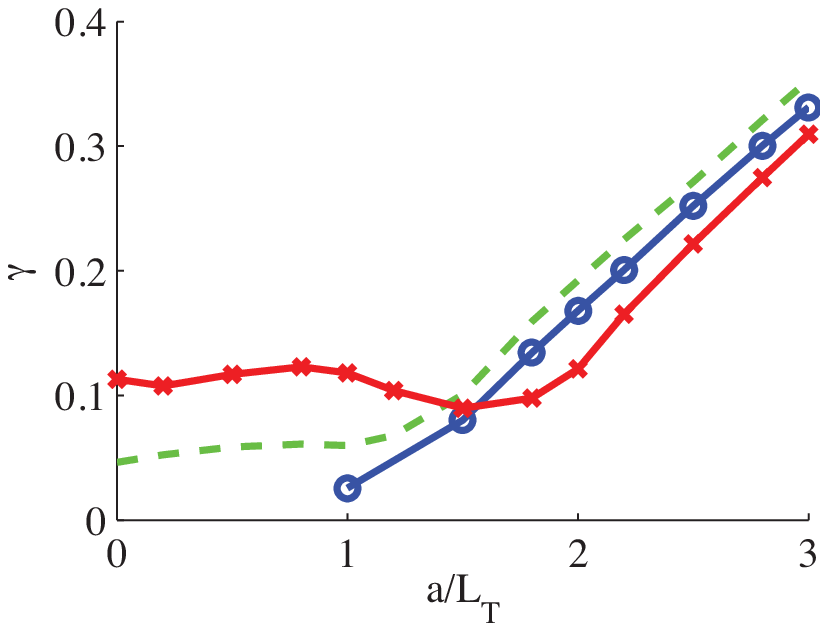}
\caption{Electrostatic ITG-TEM kinetic electron growth rates \vs temperature gradient for NCSX $\beta_{equil}=0\%$ for various density gradients: $a_N/L_n=0$ (blue circles), $a_N/L_n=1$ (green dashed line), $a_N/L_n=2$ (red crosses). (color online)}
\label{fig:ncsxbeta0ITGkinetprim_varifprim}
\end{figure}

\begin{figure} \centering
\includegraphics[scale=1.0]{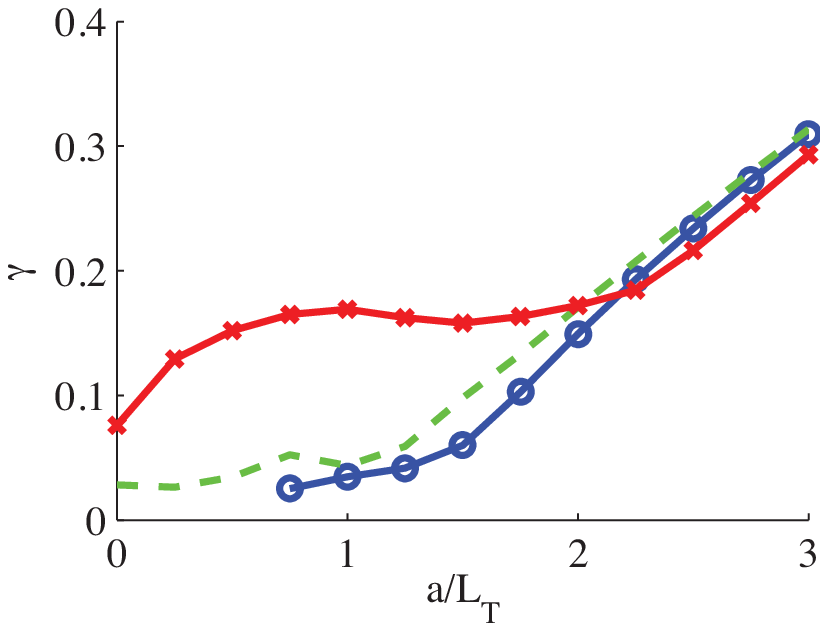}
\caption{Electrostatic ITG-TEM kinetic electron growth rates \vs temperature gradient for NCSX $\beta_{equil}=4\%$ for various density gradients: $a_N/L_n=0$ (blue circles), $a_N/L_n=1$ (green dashed line), $a_N/L_n=2$ (red crosses). (color online)}
\label{fig:ncsxbeta4ITGkinetprim_varifprim}
\end{figure}

\begin{figure} \centering
\includegraphics[scale=1.0]{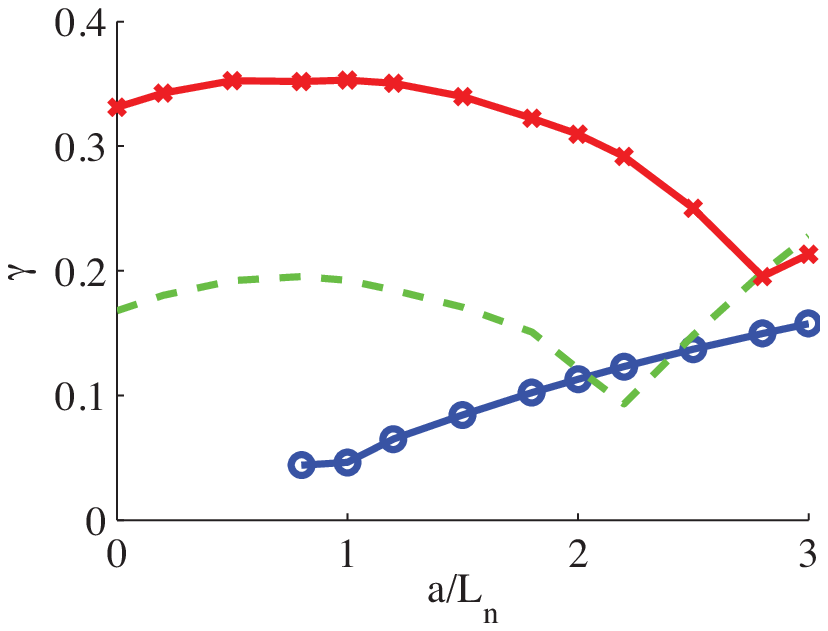}
\caption{Electrostatic ITG-TEM kinetic electron growth rates \vs density gradient for NCSX $\beta_{equil}=0\%$ for various temperature gradients: $a_N/L_T=0$ (blue circles), $a_N/L_T=2$ (green dashed line), $a_N/L_T=3$ (red crosses). (color online)}
\label{fig:ncsxbeta0ITGkinefprim_varitprim}
\end{figure}

\begin{figure} \centering
\includegraphics[scale=1.0]{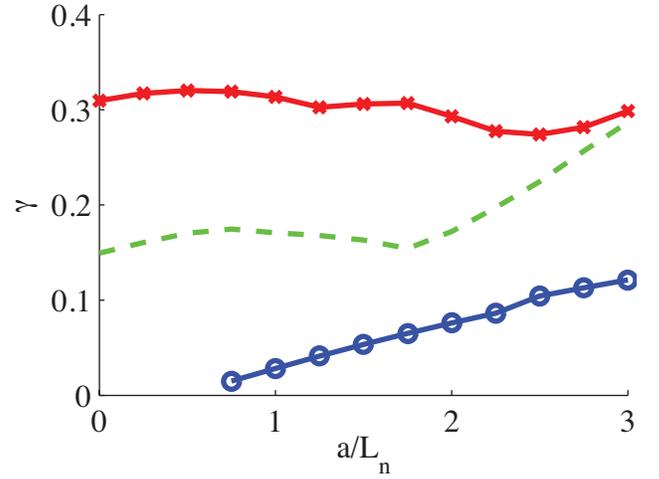}
\caption{Electrostatic ITG-TEM kinetic electron growth rates \vs density gradient for NCSX $\beta_{equil}=4\%$ for various temperature gradients: $a_N/L_T=0$ (blue circles), $a_N/L_T=2$ (green dashed line), $a_N/L_T=3$ (red crosses). (color online)}
\label{fig:ncsxbeta4ITGkinefprim_varitprim}
\end{figure}

Figures \ref{fig:ncsxbeta0ITGkinetprim_varifprim}-\ref{fig:ncsxbeta4ITGkinetprim_varifprim} show growth rates \vs $a_N/L_T$ (where $a_N/L_T=a_N/L_{Ti}=a_N/L_{Te}$) for several values of $a_N/L_n$, for both the equilibrium with $\beta=0\%$ and that with $\beta=4\%$, for the $k_y\rho_i\in[0.4, 1.8]$ with the highest growth rate. Both have the same general trend: for  all values of $a_N/L_n$, past a critical temperature gradient, the growth rates increase almost linearly with $a_N/L_T$, indicating that this mode is driven by the temperature gradient. When $a_N/L_n=0$, in both cases, there appears to be a critical temperature gradient, which is lower than in the adiabatic electron case. Here, $a_N/L_{T,crit,\beta=0}\approx0.75$ and $a_N/L_{T,crit,\beta=4}\approx0.25$. Also as in the adiabatic case, increasing $a_N/L_n$ first further destabilizes the mode--the large linear growth begins for a lower temperature gradient than for $a_N/L_n=0$--but then it is stabilizing for higher density gradients (this is more easily seen in Figures \ref{fig:ncsxbeta0ITGkinefprim_varitprim}-\ref{fig:ncsxbeta4ITGkinefprim_varitprim}). Though, as density gradient increases, the value of the flat part of the growth rate for low $a_N/L_T$ increases: this is a density-gradient-driven regime. Comparing the two $\beta$ equilibria, the $\beta=4\%$ growth rates for the flat part of the plot are lower than the $\beta=0\%$ case, for $a_N/L_n=0,1$, and higher for $a_N/L_n=2$. It appears that the ``critical gradients,'' for the strong linear growth at higher $a_N/L_T$, are lower, as well as the critical gradient for $a_N/L_n=0$. 

These results seem to differ some from Ref. \onlinecite{ernst_role_2009}, which appears to find a larger region of stability for sufficiently small $a_N/L_n$ and $a_N/L_T$, for a particular set of tokamak parameters.  Ref. \onlinecite{ernst_role_2009}, however, included finite collisions, while these studies are collisionless. Including finite collisions in these simulations may increase the stability window for TEM at weak gradients, as has been found in tokamaks \cite{ernst_role_2006} and STs.\cite{granstedt_thesis_2012}

Figures \ref{fig:ncsxbeta0ITGkinefprim_varitprim}-\ref{fig:ncsxbeta4ITGkinefprim_varitprim} show growth rates \vs $a_N/L_n$ for several values of $a_N/L_T$ (again for both equilibria and for the $k_y\rho_i\in[0.8, 1.4]$ with the highest growth rate). Here, for $a_N/L_T>1$, one can more easily see the increased destabilization of the mode as $a_N/L_n$ increases, until about $a_N/L_n=1$, when the growth rate decreases. For values of the temperature gradient lower than the adiabatic electron critical temperature gradient of $a_N/L_T\approx1.3$, the mode is density-gradient driven: the growth rate increases slowly with $a_N/L_n$. The growth rates are again higher for the $\beta=0$ case.
  
\subsection{Electromagnetic simulations}

\begin{figure} \centering
\includegraphics[scale=1.0]{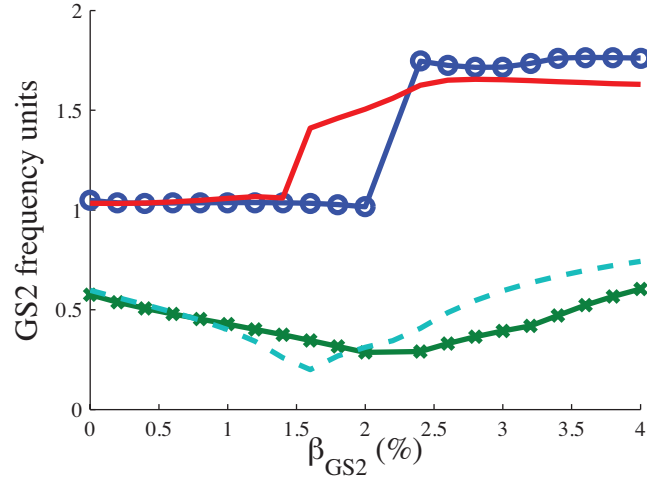}
\caption{Growth rates and real frequencies, in normalized units $(\gamma_{GS2},\omega_{GS2})=(\gamma,\omega)/(v_T/a)$ as a function of $\beta_{GS2}$, for equilibrium $\beta=0\%$ ($\omega$: red solid line, $\gamma$: blue dashed line) and $\beta=4\%$ ($\omega$: blue circles, $\gamma$: green crosses).  $k_y\rho_i=1.0$. (color online)}
\label{fig:beta04}
\end{figure}

As a preliminary investigation of electromagnetic effects, the GS2 beta parameter, $\beta_{input}$, was scaled using a fixed equilibrium, with temperature gradient $a_N/L_T=5$. For all following discussions, the notation used is $\beta_{GS2}=2\beta_{input}$, to convert from $\beta$ for a single species to percent $\beta$ for two species (an electron and an ion species of equal $T$ and $n$). In order to demonstrate the effect that instability-driven current fluctuations have on the growth rate, equilibrium $\beta$ was held fixed and $\beta_{GS2}$ scanned. Figure \ref{fig:beta04} compares two $\beta_{GS2}$ scans based on configurations with equilibrium $\beta=0\%$ and $4\%$. The frequencies and growth rates match closely when $\beta_{GS2}=0\%$. But, the apparent mode switch occurs earlier for the $\beta=0\%$ equilibrium (around $\beta_{GS2}=1.5\%$) than the $\beta=4\%$ equilibrium ($\beta_{GS2}=2.0\%$). In addition, the values in growth rate and frequency differ by about $20\%$ when $\beta_{GS2}=4\%$, indicating that matching this GS2 parameter with the equilibrium value does matter. The general trend, similar to tokamak results, is that $\beta_{GS2}$ is stabilizing to the ITG mode at moderate values, but the fastest growing mode switches character to a high frequency mode (perhaps a kinetic ballooning mode) at higher $\beta_{GS2}$. Equilibrium $\beta$ is stabilizing for this higher frequency instability (this is the stabilizing mechanism that can give rise to the second stability regime for MHD ballooning modes \cite{wesson_tokamaks_1997}).

\begin{figure} \centering
\includegraphics[scale=1.0]{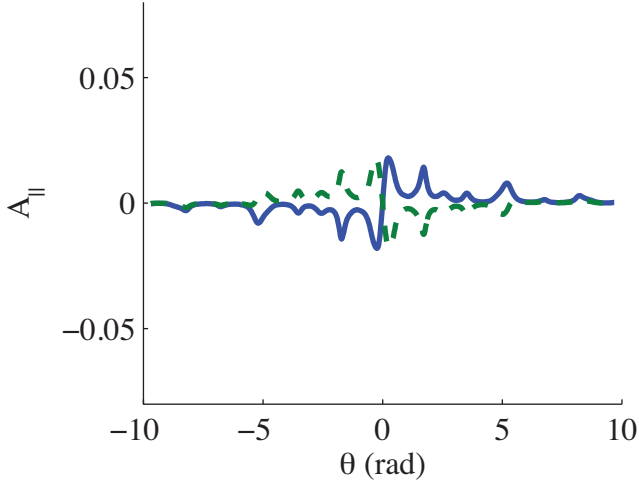}
\caption{$A_{||}$ for $\beta_{equil}=4\%$, $\beta_{GS2}=1\%$, $k_y\rho_i=1.0$. Blue: $Im(A_{||})$, Green: $Re(A_{||})$. (color online)}
\label{fig:beta4APARefcn1}
\end{figure}

\begin{figure} \centering
\includegraphics[scale=1.0]{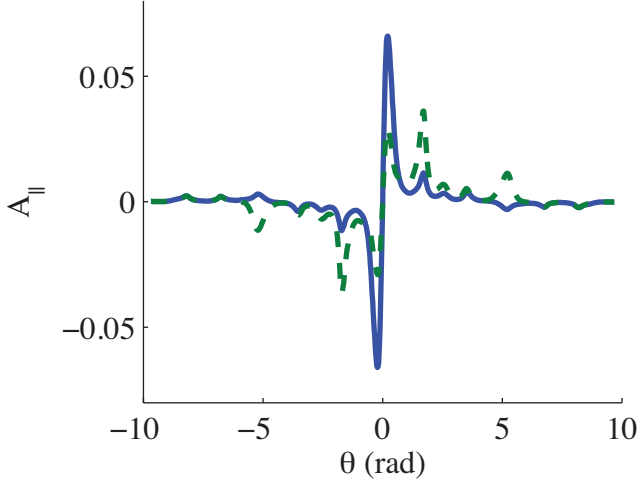}
\caption{$A_{||}$ for $\beta_{equil}=4\%$, $\beta_{GS2}=4\%$, $k_y\rho_i=1.0$. Blue: $Im(A_{||})$, Green: $Re(A_{||})$. (color online)}
\label{fig:beta4APARefcn4}
\end{figure}

\begin{figure} \centering
\includegraphics[scale=1.0]{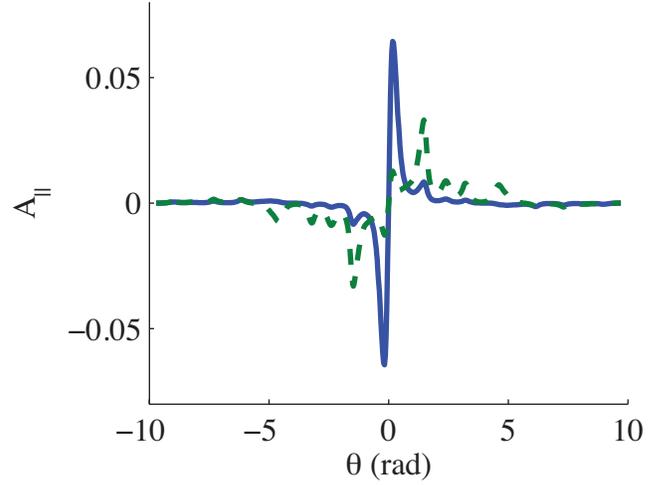}
\caption{$A_{||}$ for $\beta_{equil}=0\%$, $\beta_{GS2}=4\%$, $k_y\rho_i=1.0$. Blue: $Im(A_{||})$, Green: $Re(A_{||})$. (color online)}
\label{fig:beta0APARefcn4}
\end{figure}

Figures \ref{fig:beta4APARefcn1}-\ref{fig:beta4APARefcn4} show the magnetic potential, $A_{||}$, for $\beta_{equil}=4\%$ and $\beta_{GS2}=1\%$ and $4\%$. This demonstrates the effect $\beta_{GS2}$ has on the perturbed magnetic fields--the magnitude of $A_{||}$ is much larger in the $\beta_{GS2}=4\%$ case than in the $\beta_{GS2}=1\%$ case. Figure \ref{fig:beta0APARefcn4} is the magnetic potential,  $A_{||}$, for $\beta_{equil}=0\%$ and $\beta_{GS2}=4\%$. Note that it is almost identical to Figure \ref{fig:beta4APARefcn4},  $\beta_{equil}=4\%$ and $\beta_{GS2}=4\%$, demonstrating that only the $\beta_{GS2}$ parameter affects the fluctuating magnetic fields, not $\beta_{equil}$.


\begin{figure} \centering
\includegraphics[scale=1.0]{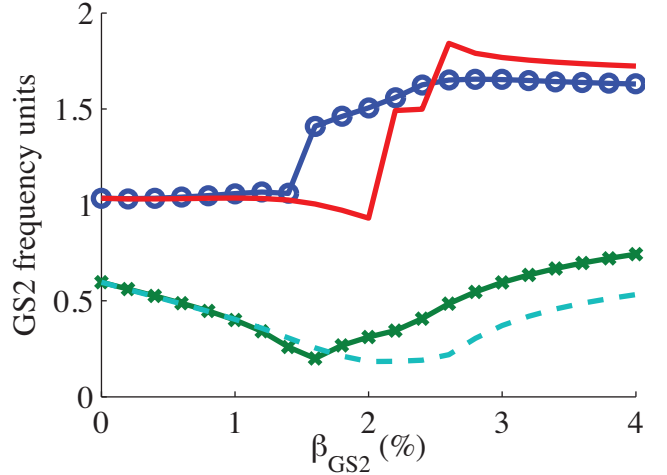}
\caption{NCSX $\beta_{equil}=0\%$: growth rates and frequencies ($(\gamma_{GS2},\omega_{GS2})=(\gamma,\omega)/(v_T/a)$) \vs $\beta_{GS2}$ for $\delta B_{||}\neq0$ ($\omega$: blue circles, $\gamma$: green crosses) and $\delta B_{||}=0$ ($\omega$: red solid line, $\gamma$: blue dashed line). (color online)}
\label{fig:beta0faperp01}
\end{figure}

\begin{figure} \centering
\includegraphics[scale=1.0]{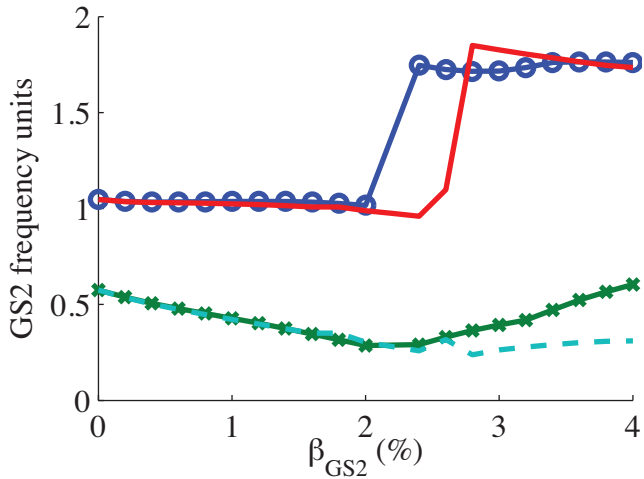}
\caption{NCSX $\beta_{equil}=4\%$: growth rates and frequencies ($(\gamma_{GS2},\omega_{GS2})=(\gamma,\omega)/(v_T/a)$) \vs $\beta_{GS2}$ for $\delta B_{||}\neq0$ ($\omega$: blue circles, $\gamma$: green crosses) and $\delta B_{||}=0$ ($\omega$: red solid line, $\gamma$: blue dashed line). (color online)}
\label{fig:beta4faperp01}
\end{figure}

In electromagnetic GS2 runs, one always includes $\delta B_{\perp}=\nabla A_{||}\times\hat{z}$ when calculating the fields, but one can choose to include $\delta B_{||}=\nabla_{\perp}\times\mathbf{A}_{\perp}$\cite{howes_astrophysical_2006} or set it to zero. One might want to ignore this term to save computational time. Figures \ref{fig:beta0faperp01}-\ref{fig:beta4faperp01} demonstrate the importance of including $\delta B_{||}$ for high $\beta_{GS2}$ values. For $\beta_{GS2}\lesssim1.5\%$, the growth rates and frequencies for $\delta B_{||}=0$ and $\delta B_{||}\neq0$ are approximately equal, because $\beta_{GS2}$ scales the $\delta B_{||}$ field, so that when $\beta_{GS2}$ is low, the contribution from $\delta B_{||}$ is small. However, as $\beta_{GS2}$ increases past $\beta_{GS2}=2\%$, the contribution from $\delta B_{||}$ increases: including $\delta B_{||}$ has a destabilizing effect at higher $\beta_{GS2}$.

\section{Conclusion}

Understanding the effects of stellarator geometry on gyrokinetic turbulence, along with how much of an effect turbulence has on confinement in current experiments, is of paramount importance for designing future magnetic condiment fusion devices. Drift-wave instabilities are believed to cause turbulence, and can be modeled using several gyrokinetic codes, including GS2. The nonlinear gyrokinetic turbulence code GS2's non-axisymmetric geometry capabilities were linearly benchmarked for an NCSX equilibrium with GENE and GKV-X. The growth rates and real frequencies of an adiabatic ITG mode agreed to within $8\%$ for all three codes. Coupled with a previous GS2 benchmark with FULL,\cite{baumgaertel_simulating_2011} all four gyrokinetic stellarator codes have been benchmarked successfully against each other.

Extensive studies of instabilities in two NCSX equilibria were conducted. Comparing NCSX equilibria of differing $\beta$ values revealed that the $\beta=4\%$ equilibrium was marginally more stable than the $\beta=0\%$ equilibrium in the adiabatic-electron ITG mode case, but less stable in kinetic electron ITG-TEM mode case. However, their electrostatic adiabatic ITG mode and electrostatic kinetic ITG-TEM mode growth rate dependencies on $a_N/L_T$, $a_N/L_n$, and $k_y\rho_i$ were similar.

There are two effects of finite plasma $\beta$ on microinstabilities. The first is created by the changes in magnetic geometry, and this affects even electrostatic modes. The second effect is due to fluctuating currents (and magnetic fields), and this can be varied as an independent parameter in calculations (although not in the real world). It was demonstrated through the electromagnetic ITG-TEM modes that $\beta_{GS2}$ must be set consistently with the equilibrium $\beta$ in order to have physical results. For a fixed magnetic equilibrium, the effect of $\beta_{GS2}$ on magnetic fluctuations is at first stabilizing (from $\beta_{GS2}=0\%-2\%$) and then destabilizing (for $\beta_{GS2}\approx2\%-4\%$).  It is important to keep the parallel component of magnetic fluctuations $\delta B_{||}$ for $\beta_{GS2}>2\%$.

Future work includes studying other stellarator configurations and comparing stellarator and tokamak equilibria for linear gyrokinetic instability. In addition, GS2 is fully capable of nonlinear stellarator simulations, and ultimately, one wishes to compare nonlinear turbulent fluxes with experimental measurements. GS2 will be a good tool for such use. The successful benchmarks presented here increases the confidence in the stellarator capabilities of all of the involved gyrokinetic codes.

\section{Acknowledgements}

The authors wish to thank Neil Pomphrey for producing a series of NCSX equilibria that are invaluable for geometrical parameter scans, and W. Dorland, M. A. Barnes, and W. Guttenfelder for their help with GS2. They are also grateful to Paul Bradley for encouraging the completion of this work while at LANL.  
This work was supported by the U.S. Department of Energy through the SciDAC Center for the Study of Plasma Microturbulence, the Princeton Plasma Physics Laboratory under DOE Contract No. DE-AC02-09CH11466, and Los Alamos National Security, LLC under DOE Contract No. DE-AC52-06NA25396.

\end{document}